\documentclass[12pt]{article}

\usepackage{epsfig}
\usepackage{graphics}
\usepackage{latexsym}
\usepackage{amssymb}
\usepackage{amsmath}
\usepackage{amsthm}

\newcommand{\ben}{\begin{displaymath}}
\newcommand{\een}{\end{displaymath}}
\def\one{{\sf 1}\mkern-5.0mu{\rm I}}
\def\a{\alpha}
\def\b{\beta}
\def\d{{\rm d}}
\def\g{\gamma}

\def\e{\epsilon}

\def\Spin{\mathrm{Spin}}

\def\dim{\mathrm{dim}}

\newtheorem{lemma}{Lemma}
\newtheorem{theorem}[lemma]{Theorem}

\title{Zero Energy States of Reduced Super Yang-Mills Theories in $d+1 =4,6$ and
$10$ dimensions are necessarily $\mathrm{Spin}(d)$ invariant}
\author{D. Hasler${}^{(a)}$,  J.
Hoppe${}^{(b)}$ \\ 
\vspace*{-0.05truein} \\
\normalsize\it ${}^{(a)}$ Theoretische Physik,
ETH-H\"onggerberg, CH--8093 Z\"urich\\
\normalsize\it   ${}^{(b)}$ Department of Mathematics, KTH, S-10044 Stockholm }

\begin{document}

\maketitle
\vspace{0.4cm}
\begin{abstract}
We consider reduced Super Yang-Mills Theory
in $d+1$ dimensions, 
where $d=2,3,5,9$. We present commutators to prove that for $d=3,5$
and $9$ a possible ground state 
must be a $\mathrm{Spin}(d)$ singlet. We also discuss the case $d=2$, where we
give an upper bound on the total angular momentum and show that for
odd dimensional gauge group no $\mathrm{Spin}(d)$ invariant state
exists in the Hilbert space.
\end{abstract}

\section{Introduction}

We consider models, which are obtained by dimensional reduction of
Super Yang-Mills theory with gauge group $SU(N)$ in
$d+1$ dimensions, where $d=2,3,5,9$. These models were used to formulate a
quantum theory of supermembranes living in $d+2$ dimensions, and for
$d=9$ they describe $N$ interacting $D0$ branes. It is interesting
to know, whether these models admit a possible 
zero energy  state and what the properties of such a state are. The
general belief, 
partially proven, is that for $d=2,3,5$ no zero energy state exists
and that for 
$d=9$ there exists a unique ground state.

Let us start with a very simple argument that zero energy states in
$d=9$ are $\mathrm{Spin}(d)$ invariant:
it is well known \cite{malshevan:tra,weissnicht,hop:mem} that the
supercharges, $Q_{\b}$, of reduced Yang-Mills theory (for definitions and conventions, see the
next section) 
and  
\ben
\widetilde{Q}_{\b} = m \sum_{i=1}^3 q_{iA} (\gamma^{123} \g^i)_{\b \a}
\Theta_{\a A} - \frac{m}{2} \sum_{\mu =4}^9 q_{\mu A} (\gamma^{123} \g^{\mu}
)_{\b \a} \Theta_{\a A} \ ,
\een
$\g^{123} = \g^1 \g^2 \g^3$, satisfy anti commutation relations of the
form
\ben
\{ Q_{\beta} + \widetilde{Q}_{\beta} , Q_{\beta'} + \widetilde{Q}_{\beta'} \} =
\delta_{\beta \beta'} \widehat{H} + m J_{ij}(\gamma^{123} \gamma^{ij})_{\beta
\beta'} - \frac{m}{2} J_{\mu \nu}(\gamma^{123} \gamma^{\mu \nu} )_{\beta
\beta'} + 2 q_{tA} \gamma^t_{\beta \beta'} J_A ,
\een
where $J_A , J_{ij} , J_{\mu \nu}$ are the generators of $SU(N)$, $\Spin(3)$,
$\Spin(6)$ respectively and $\widehat{H}$ is an operator, whose form is not
important here. As
\ben
\{ \widetilde{Q}_{\b} , \widetilde{Q}_{\b'} \} = \delta_{\b \b'} ( m^2
\sum_{iA} q_{iA}^2 + \frac{m^2}{4} \sum_{\mu A} q^2_{\mu A} ) =:
\left( \delta_{\beta
\beta'} \widetilde{H} \right) \, ,
\een
it immediately follows that
\begin{eqnarray} \label{eq:jens5}
\{ Q_{\b} , \widetilde{Q}_{\beta'} \} + \{ \widetilde{Q}_{\b} , Q_{\beta'} \}
& = & \delta_{\b \b'} ( \widehat{H} - H - \widetilde{H} ) 
 \\ \nonumber & &  + m  J_{ij} (\gamma^{123} \gamma^{ij})_{\beta \beta'} - 
\frac{m}{2} J_{\mu \nu}
(\gamma^{123} \gamma^{\mu \nu})_{\beta \beta'} \, ,
\end{eqnarray}
so that for $SU(N)$ invariant zero energy states $\phi, \psi$, i.e. states
annihilated by the $Q_{\b}$ and $J_A$,
\ben 
(\phi , J_{ij} \psi ) = 0 ,  \ 
  ( \phi , J_{\mu \nu} \psi ) = 0 .
\een
(just multiply (\ref{eq:jens5}) by $(\gamma^{123} \gamma^{ij})_{\beta
\beta'}$, respectively $(\gamma^{123} \gamma^{\mu \nu} )_{\b \b'}$ and sum
over $\beta$ and $ \beta'$); hence 
\ben
J_{ij} \psi = 0 , \  J_{\mu \nu} \psi = 0  ,
\een
by choosing $\phi = J_{ij} \psi$ , respectively $J_{\mu \nu} \psi$. As
(123) may be replaced by any other triple $(stu)$, $J_{st} 
\psi = 0$, for all $s,t=1,...9$, provided $Q_{\beta} \psi = 0,  J_{A}
\psi= 0$.

In the next section we will treat $d=2,3,5,9$ on equal footing and,
similar to \cite{setste:inv}, look for anti-commutators to prove that zero-energy states
have to be invariant under $\mathrm{Spin}(d)$. We do find such
anti-commutators for $d=3,5$ and $9$. For $d=2$, we give an upper bound
on the total angular momentum and show that if $SU(N)$ is odd
dimensional, i.e. $N$ even, no $\mathrm{Spin}(d)$ invariant state
exists in the Hilbert space. The discussion below generalizes to other
gauge groups.

\section{Model and Results }

Let $d=2,3,5,9$, and let $(\gamma^i)_{\a \b}$ denote the real irreducible representation of
smallest dimension, called $s_d$, of the $\g$-matrices in $d$
dimensions, i.e. the relations
$\{ \gamma^s , \gamma^t \} = 2 \delta^{st} \one$.
We have  $s_d=2,4,8,16$. The model, which we are discussing, contains the self
adjoint bosonic
degrees of freedom $q_{sA}$, $p_{sA}$ ($s=1,...,d, \, 
A=1,...,N^2-1$) and the self adjoint fermionic degrees of freedom $\Theta_{\a A}$
($\a=1,...,s_d, \, A=1,...,N^2-1$) satisfying 
\begin{eqnarray} \label{eq:bos}
[q_{sA}, p_{tB} ] & = & i  \delta_{st} \delta_{AB} , \\  \label{eq:fer}
\{ \Theta_{\a A} , \Theta_{\b B} \} & = & \delta_{\a \b} \delta_{A B}, \\
\nonumber
[q_{sA}, \Theta_{\a B}] & = & [p_{sA}, \Theta_{\a B}] = 0 .
\end{eqnarray}
More precisely, we consider the Schr\"odinger representation ($p_{sA} =
- i \partial_{sA}$) of (\ref{eq:bos}) on the Hilbert space
\ben
\mathcal{H} = L^2(\mathbb{R}^{d(N^2-1)}) \otimes \mathcal{F} ,
\een
where $\mathcal{F} \cong (\mathbb{C}^2)^{(s_d/2)(N^2-1)}$ is the
irreducible representation space of (\ref{eq:fer}). The infinitesimal
generators of the gauge group $SU(N)$ read
\ben
J_{A} =  - i f_{ABC} ( q_{tB} \partial_{tC} + \frac{1}{2} \Theta_{\a B}
\Theta_{\a C} ) ,
\een
where $f_{ABC}$ are real, antisymmetric structure constants of $SU(N)$.
The physical Hilbert space $\mathcal{H}_{\mathrm{phys}}$, given by
the $SU(N)$ invariant states in $\mathcal{H}$, is the Hilbert space of
the model.
We have a representation of $\Spin(d)$ on $\mathcal{H}$ ($\mathcal{H}_{\mathrm{phys}}$), with
infinitesimal generators 
\begin{eqnarray*}
J_{st} & = & - i ( q_{sA} \partial_{tA} - q_{tA} \partial_{sA} ) - \frac{i}{4}
\Theta_{\a A} \g^{st}_{\a \b} \Theta_{\b A} \\
& \equiv & L_{st} + M_{st}  \ ,
\end{eqnarray*}
where $\g^{st} = \frac{1}{2} [\gamma^s, \gamma^t ]$.
The supercharges are given by
\ben
Q_{\b} = \Theta_{\a A} ( - i\g^t_{\a \b} \partial_{t A} + \frac{1}{2} f_{ABC}
q_{sB} q_{tC} \g^{st}_{\b \a} ) \ ,
\een
and the Hamiltonian by
\ben
H = - \Delta + \frac{1}{2} f_{ABC}q_{sB}q_{tC} f_{ADE} q_{sD} q_{tE} +
i q_{sA} f_{ABC} \Theta_{\a B} \Theta_{\b C} \g^{s}_{\a \b} \ .
\een
The anti-commutation relations for the supercharges are
\ben
\{ Q_{\a} , Q_{\b} \} = \delta_{\a \b} H + 2 \g^s_{\a \b} q_{sA}J_A \ ,
\een
On $\mathcal{H}_{\mathrm{phys}}$ this reads 
\ben
\{ Q_{\a} , Q_{\b} \} = \delta_{\a \b} H . 
\een
We note
that the Operators $Q_{\a}$ and $H$ are self adjoint on their maximal domain, i.e. 
\begin{eqnarray*}
\mathcal{D}(Q_{\a} ) & = & \{ \psi \in \mathcal{H} | (Q_{\a} \psi
)_{\mathrm{dist}} \in 
\mathcal{H} \}, \\ 
\mathcal{D}(H) & = & \{ \psi \in \mathcal{H} | (H \psi )_{\mathrm{dist}} \in \mathcal{H} \} ,
\end{eqnarray*}
where $( \ \cdot \ )_{\mathrm{dist}}$ is understood in the sense of
distributions. The 
restrictions of $H$ and $Q_{\a}$ to $\mathcal{H}_{\mathrm{phys}}$ are also
self adjoint. We are only interested in $SU(N)$ invariant states,
i.e. states in 
$\mathcal{H}_{\mathrm{phys}}$. By definition $\psi$
is a zero energy state iff $\psi \in \mathcal{H}_{\mathrm{phys}} \cap
\mathrm{Ker} H$. We want to prove the following 
\begin{theorem} \ \label{th:d=2359}
\begin{description}
\item[(a)] For $d=3,5,9$, a possible zero energy state is a $\Spin (d)$
singlet.
\item[(b)] For $d=2$, a possible zero energy state $\psi$ satisfies
\ben
\| J_{st} \psi \| \leq \frac{3}{2} \cdot \dim \, SU(N) \, \| \psi \| .
\een
\end{description}
\end{theorem}
We start with 
\begin{lemma} \label{le:onc1} 
We have the following (formal) anti-commutator relations.

\begin{description}
\item[(a)] For $d=2,3,5,9$, we have with $b_{\a}^{uv} := \frac{1}{s_d}
(  q_{uA} \g^{v}_{\a \e}
\Theta_{\e A} - q_{vA} \g^{u}_{\a \e} \Theta_{\e A})$,
\ben
\left\{ Q_{\a} , b_{\a}^{uv} \right\} =   J_{uv} + (8/s_d  - 1) M_{uv}  \ . 
\een
\item[(b)] For $d=2,3,9$, we have with $c_{\a}^{uv} := \frac{1}{s_d}(
  \g^{w} \g^{uv} )_{\a \e} q_{w D} \Theta_{\e D}$,
\ben
\left\{ Q_{\a} , c_{\a}^{uv} \right\} =  J_{uv} +
(4d/s_d  - 1 ) M_{uv}
\een 
\item[(c)] For $d=3,5,9$, we have with $a_{\a}^{uv} =  \frac{1}{4d-8}  \cdot \left( (
{4d} - s_d ) b_{\a}^{uv} - ( {8} - s_d) c_{\a}^{uv} \right)$,
\ben
\{ Q_{\a} , a_{\a}^{uv} \} = J_{uv} .
\een
\end{description}
\end{lemma}

\begin{proof} 
\ \ \newline  \noindent
By a straight forward calculation we find for $d=2,3,5,9$
\begin{equation} \label{eq:lem2(1)}
\{ Q_{\a} , q_{uD} \Theta_{\e F} \} = - i \g^{u}_{\b \a} \Theta_{\b D}
\Theta_{\epsilon F} + \g^s_{\epsilon \a } q_{u D} ( - i \partial_{sF}) +
\frac{1}{2} f_{FBC} q_{uD} q_{sB} q_{tC} \g^{st}_{\a \epsilon} \ .
\end{equation}
\newline \noindent
(a) We have, using (\ref{eq:lem2(1)}),
\begin{equation} \label{eq:com1}
\{ Q_{\a} , q_{uD} \g^{v}_{\a \e} \Theta_{\e D} \} = 
- i ( \g^{u} \g^{v} )_{\b \e} \Theta_{\b D} \Theta_{\e D} + 
s_d q_{uD} ( - i \partial_{vD}) +
\frac{1}{2} f_{DBC} q_{uD} q_{sB} q_{tC} \g^{st}_{\a \e} \g^{v}_{\a \e} \ .
\end{equation}
The last term in (\ref{eq:com1}) vanishes since the trace over
the $\g$-matrices equals zero. We find 
\begin{eqnarray*}
\{ Q_{\a} , q_{uA} \g^v_{\a \e}  \Theta_{\e A} - q_{v A} \g^{u}_{\a \e}
\Theta_{\e A} \} & = & 
- i s_d ( q_{uA} \partial_{vA} - q_{vA} \partial_{uA} )
- i \Theta_{\a A} 2 \g^{uv}_{\a \e} \Theta_{\e A} \\
& = &
s_d J_{uv} + ( 8 - s_d ) M_{uv}  \ .
\end{eqnarray*}
\newline \noindent
(b) We have, using (\ref{eq:lem2(1)}),
\begin{eqnarray*}
\{ Q_{\a} , (\g^{w} \g^{uv})_{\a \epsilon} q_{wD} \Theta_{\epsilon D} \} & = &
d (-i) \g^{uv}_{\b \epsilon} \Theta_{\b D} \Theta_{\epsilon D}
+ s_d (-i) ( q_{uD} \partial_{vD} - q_{vD} \partial_{u D} )  \\
&  & - \frac{1}{2} \mathrm{Tr}( \g^w \g^{uv} \g^{st}) f_{DBC} q_{wD} q_{sB} q_{tC} \\
& = & s_d J_{uv} + ( 4d - s_d ) M_{uv} \ ,
\end{eqnarray*}
where the term in the second line is zero, as the trace over the five
$\g$-matrices vanishes. 
\newline 
\noindent
(c) follows by a linear combination of (a) and (b).
\end{proof}

We note that the action of $\Spin(d)$ leaves the kernel of $H$ invariant. Let
$\varphi , \psi \in \mathrm{Ker} H \cap
\mathcal{H}_{\mathrm{phys}}$. Then $\varphi , \psi \in 
\mathrm{Ker} Q_{\b}$ for all $\b$ and by elliptic regularity $\varphi,\psi \in
C^{\infty}$. We assume 
that $\psi$ lies in an irreducible representation space of $\Spin(d)$. Hence
$J_{uv} \psi \in \mathrm{Ker} H \cap \mathcal{H}_{\mathrm{phys}}$. Let
$d=3,5,9$. By Lemma \ref{le:onc1} (c), we have 
\ben
Q_{\a} a^{uv}_{\a} \psi = \{ Q_{\a} , a^{uv}_{\a} \} \psi =  J_{uv}
\psi \ \in \mathcal{H} \ . 
\een 
Taking the scalar product with $\varphi$, we want to bring $Q_{\a}$ to
the other side, i.e. integrate by parts. Therefore we regularize as in
\cite{setste:inv}.
There exists a function
$\chi : \ [0, \infty)  \to \mathbb{R}$ in $C^{\infty}$,
such that
\ben
\chi(r) = \left\{ 
\begin{array}{ll} 
1 & r \leq 1 \\
\in [0,1] & 1 < r < 3 \\
0 & 3 \leq r 
\end{array} \right.,
\een
and $| \chi'(r) | \leq 1$. Define $g_n(q) \equiv \chi(|q|/n)$. By dominated
convergence,
\begin{eqnarray} 
( \varphi , J_{uv}  \psi )  & = & 
\lim_{n \to \infty} ( \varphi , g_n Q_{\a} a^{uv}_{\a} \psi ) \nonumber \\
& = & \lim_{n \to \infty} ( \varphi , [ g_n , Q_{\a} ] a^{uv}_{\a} \psi ) +
 \lim_{n \to \infty} ( \varphi , Q_{\a} g_n a^{uv}_{\a} \psi )   \ . \label{eq:byd2}
\end{eqnarray}
The second term in (\ref{eq:byd2}) vanishes since $g_n a_{\a}^{uv}
\psi \in C^{\infty}_0$ is in the domain of $Q_{\a}$ and $Q_{\a}$ is
self adjoint. By 
\ben
|[Q_{\b} , g_n ] |_{\mathcal{F}} \leq \mathrm{const.} \cdot \frac{1}{n}
\chi'({|q|}/{n}) ,
\een
where $|  \cdot |_{\mathcal{F}}$ stands for the the norm in $\mathcal{F}$ or
the operator norm in 
$\mathcal{L}(\mathcal{F})$, the first term in (\ref{eq:byd2}) vanishes
using the 
following estimate.
\begin{eqnarray*}
|( \varphi , [ g_n , Q_{\a} ] a^{uv}_{\a} \psi ) | \leq 
\mathrm{const.}   \cdot \int_{n \leq |q| \leq 3n}{ n \cdot \frac{1}{n}
| \varphi|_{\mathcal{F}} 
| \psi|_{\mathcal{F}} \d q} \ \rightarrow 0 \qquad \mathrm{for} \ \ n
\rightarrow \infty . 
\end{eqnarray*}
Hence $(\varphi , J_{uv} \psi ) = 0$.
Choosing $\varphi = J_{uv} \psi $, we find
\ben
( J_{uv} \psi , J_{uv} \psi ) = 0 .
\een
By linear combination, it follows that for $d=3,5,9$ all states in
$\mathrm{Ker}H \cap \mathcal{H}_{\mathrm{phys}}$ are
$\Spin(d)$ singlets. For $d=2$,  
we use Lemma
\ref{le:onc1} (a) or (b) and find by an analogous argument $(\varphi,
(J_{st} + 6M_{st} ) \psi)=0.$ Choosing $\varphi = J_{st} \psi$,
we obtain
\begin{eqnarray*}
(J_{12} \psi , J_{12} \psi ) & \leq & 6 | ( J_{12} \psi , M_{12} \psi ) | \\
& \leq & 6 \| J_{12} \psi \| \cdot \| M_{12} \psi \| \ .
\end{eqnarray*}
A real irreducible representation of the $\g$-matrices in 2 dimensions is
given by $\g^1 = \sigma^1 , \, \g^2 =  - \sigma^3$. 
In this representation we have $\g^{12} = \frac{1}{2} [\g^1, \g^2 ]
= i \sigma^2$. It follows that
\begin{eqnarray*}
\| J_{12} \psi \| & \leq & 6 \| M_{12} \psi \| \\
& = & 6 \| \frac{i}{4} \Theta_{\a A} \g^{12}_{\a \b}
\Theta_{\b A} \psi \|  \\
& = & 6 \| \frac{i}{2} \Theta_{1 A} \Theta_{2 A} \psi \| \\
& \leq & \frac{3}{2} \dim \, SU(N)\| \psi \| .
\end{eqnarray*}
By linear combination the above equation holds for all states in
$\mathrm{Ker}H \cap \mathcal{H}_{\mathrm{phys}}$. Hence Theorem \ref{th:d=2359} follows.

The case $d=2$ is special as the following theorem shows. 
\begin{theorem}  \label{spindd=2}
For $d=2$ and odd dimensional gauge
group $SU(N)$ no $\mathrm{Spin}(d)$ invariant state exists in
$\mathcal{H}$.
\end{theorem}

\begin{proof}
By definition
\ben
J_{12} = - i( q_{1A} \partial_{2A} -  q_{2A} \partial_{1A}) - \frac{i}{4}
\Theta_{\a A} \g^{12}_{\a \b} \Theta_{\b A} .
\een
As above, we choose $\g^1 = \sigma^1 , \, \g^2 =  - \sigma^3$.
We define the following annihilation and creation operators
\ben
 \frac{\partial}{\partial \lambda_A} = \frac{1}{\sqrt{2}}( \Theta_{1A}
 + i \Theta_{2A} ) , \qquad 
\lambda_A = \frac{1}{\sqrt{2}}( \Theta_{1A}-  i
\Theta_{2A} ) .
\een
We find
\ben
J_{12} = L_{12} - \frac{i}{2} \Theta_{1A} \Theta_{2A} =
L_{12} - \frac{1}{2}\lambda_A \frac{\partial}{\partial \lambda_A} +
\frac{1}{4} \cdot \mathrm{dim} \, SU(N) . 
\een
Assume $\psi$ is $\mathrm{Spin}(d)$-invariant, i.e. $J_{12} \psi = 0$. Then
\ben
\left( L_{12} - \frac{1}{2}
\lambda_A \frac{\partial}{\partial \lambda_A} \right) \psi =  - \frac{1}{4}
\cdot 
\mathrm{dim}\, SU(N) \psi .
\een
If $\mathrm{dim} \, SU(N)$ is odd this contradicts that the spectrum of
$L_{12} - \frac{1}{2} 
\lambda_A \frac{\partial}{\partial \lambda_A }$ only takes values in
$\frac{1}{2} 
\mathbb{Z}$. Hence the claim follows.
\end{proof}

{\bf Acknowledgments.\/} We thank J. Fr\"ohlich and G.M. Graf for
useful discussions, and S. Sethi for pointing out to us reference
\cite{setste:inv}.

\end{document}